# Ground State Phase Diagrams and Magnetic Properties of the Double Perovskite $Pb_2FeReO_6$

S. MTOUGUI[1], S. IDRISSI[1], H. LABRIM[2], N. EL MEKKAOUI[1], I. EL HOUSNI[1], S. ZITI[3], R. KHALLADI[1] and L. BAHMAD[1,*]

[1] Laboratoire de la Matière Condensée et des Sciences Interdisciplinaires (LaMCScI), Mohammed V University of Rabat, Faculty of Sciences, B.P. 1014 Rabat, Morocco.

[2] USM/DERS/Centre National de l'Energie, des Sciences et des Techniques Nucléaires (CNESTEN), Rabat, Morocco.

[3] Intelligence Artificial and Security of Systems, Mohammed V University of Rabat, Faculty of Sciences, B.P. 1014 Rabat, Morocco.

**Abstract:**

The Half-Metallic Ferromagnetic behavior of some double perovskites attracts much interest. In this work, we studied the magnetic behavior of the double perovskite $Pb_2FeReO_6$. The magnetic atoms are Fe and Re and have the spins $S=5/2$ and $\sigma=1$, respectively. In a first step, we provide a theoretical study of the ground state phase diagrams .In fact, we present and discuss the stable configurations from the all 6x3=18 possible configurations. Secondly, the magnetic properties of this compound when varying different physical parameters is carried out. Besides, we used Monte Carlo simulations (MCS), under the Metropolis algorithm to provide the magnetic behavior of the studied system as a function of the temperature, the crystal field, the exchange coupling interactions and the external magnetic field. In addition, we studied and discussed the critical temperature of the double perovskite $Pb_2FeReO_6$. To complete our study, we presented and analyzed the hysteresis loops for specific values of physical parameters.

[*]) Corresponding authors: bahmad@fsr.ac.ma (L.B.); sara.mtougui@gmail.com (S.M.)

## I. Introduction

Recently, the perovskites with mixed-metal oxides have attracted much interest for industrial and photovoltaic applications because of their low price, adaptability, and thermal stability. Such materials have shown interesting magnetic and optical properties [1].

During the years 1998 and 1999, Kobayashi *et al.* [2, 3] discovered the half-metallic ferromagnetic (HM-FM) properties, the large tunneling magnetoresistance (TMR) and high Curie temperature (Tc) for the compounds $Sr_2FeMoO_6$ and $Sr_2FeReO_6$, respectively. The HM-FM behavior has also been detected in many double perovskites such as $Ho_2NiMnO_6$ [4], $Ba_2CrMoO_6$ and $Ba_2FeMoO_6$ [5]. Additionally, these compounds present 100% spin polarization of the conduction electrons at the Fermi level and contribute to the development of technological applications in the devices of single-spin electron source and high-efficiency magnetic sensors [6, 7]. Moreover, recent works have been motivated the study of the double perovskite compounds with a general chemical formula $A_2MM'O_6$; where A is an alkaline-earth metal atom or rare earth metal atom, M stand for the 3d transition-metal (TM) atom and M' is the 4d/5d TM atom [8]. The majority of the double perovskites have been found to take a rock-salt crystal structure alternating perovskite units $AMO_3$ and $AM'O_3$ along three crystallo-graphical axes. The corners of each perovskite unitary alternately are occupied by the TM atoms M and M'. The oxygen atoms are located between M and M' atoms forming alternate $MO_6$ and $M'O_6$ octahedra. The large alkaline-earth-metal atom or rare-earth-metal atom A occupies the body-centered site with a (12-fold) oxygen coordination in each unit [9].

The combinations of the atoms A, M and M' can be chosen depending on the desired applications. For spintronic devices $Sr_2FeMoO_6$ [2] and $Sr_2FeReO_6$ [3] represent good examples, for multiferroicity to quote the $Ba_2NiMnO_6$ [10], in magneto-dielectric materials $La_2NiMnO_6$ [11, 12] and in magneto optic applications $Sr_2CrReO_6$ and $Sr_2CrOsO_6$ [13].

During the year 2009, Nishimura *et al.* [14] elaborated a new interesting double perovskite, $Pb_2FeReO_6$. Earlier, in 2001, Wolf *et al.* [12] presented the importance of thin films of $Pb_2FeReO_6$ sandwiching a thin nonmagnetic film (Cu film in spin valve device) or a very thin insulating layer ($Al_2O_3$ film in magnetic tunnel junction device). So maintaining the HM-FM behavior of the $Pb_2FeReO_6$ thin films is interesting for technological applications in magneto resistive and spintronic devices [9].

Experimentally, the double perovskite $Pb_2FeReO_6$ was prepared under the special conditions of pressure 6 GPa and temperature 1000 °C. The synchrotron X-ray powder diffraction study, showed that this perovskite crystallizes in the tetragonal structure (a =b= 5.62A° and c = 7.95A°) belonging to the space group I4/m with the presence of $Pb^{2+}$ ions at the A site. At temperatures less than 23 K, no structural transition to the lower symmetry was recorded. This perovskite displayed a ferromagnetic transition at 420 K [14]. The Curie temperature (Tc) of this perovskite is about 420 K; this value is slightly higher than that one of $Sr_2FeReO_6$, which is about 401 K [15]. The distances between the four equatorial metal transitions and oxygen are less than they are in the two axial transitions. Also, this compound shows the existence of the Jahn–Teller structural distortion in $FeO_6$ and $ReO_6$ octahedra [16]. The $Fe^{3+}$ ions are in the states $3d^5$ with the magnetic moment of spin S=5/2 and $Re^{5+}$ ions are in the states $5d^2$ with the magnetic moment of spin σ=1. The magnetic moments of $Fe^{3+}$ and $Re^{5+}$ are 3.929 and -0.831$\mu_B$, respectively. Therefore, there exists an anti-ferromagnetic (AFM) coupling via oxygen between these two ions [16]. The half-metallic ferromagnetic (HM-FM) behavior exhibits a potential application of this new compound in magneto electronic and spintronic devices [16]. In one of our recent works [17], we have studied and discussed the magnetic properties of the perovskite $BiFeO_3$, showing the hysteresis loops of this compound.

Motivated by the above considerations, the aim of this paper is to study and predict the phases diagrams and the magnetic properties of the double perovskite $Pb_2FeReO_6$, using Monte Carlo simulations (MCS). In a first step, we perform the ground state phase diagrams for the perovskite $Pb_2FeReO_6$; then we explore the effect of physical parameters, such as the temperature, the exchange coupling interactions, the crystal field and the external magnetic field on the behavior of the total magnetizations and susceptibilities. In addition, the behavior of the hysteresis loops for specific values of the physical parameters has been carried out. To achieve this goal, we will provide the ground state phase diagrams and the obtained results of Monte Carlo Simulations. This method is based on computer simulations under the Metropolis algorithm, often used to solve difficult physical problems [18-22]. The geometry of the double perovskite $Pb_2FeReO_6$ has been established using the Vesta software [23]. The critical exponents of this double perovskite $Pb_2FeReO_6$ have been established using the well-known finite-size scaling functions [24–28].

In some of our recent works, we have applied both the Monte Carlo simulations and Ab-initio approach, in order to investigate and discuss the magnetic properties of some multi-ferroic alloys [29-32].

This paper is organized as following. In Section II, we present the physical model. In section III, we discuss the obtained results. We finally complete this work with a conclusion in section IV.

## II. Theoretical model

To study the behavior of such complex spin systems, we used Monte Carlo Simulations. This method showed its efficacy in the study of magnetic behavior of magnetic materials [33, 34]. The studied compound consists of two interpenetrating sub-lattices. One sublattice has spins σ and the other sublattice has spins S. The spins S have the spins σ and spins S as nearest neighbors and vice versa. The geometry of the studied system is belonging to the space group I4/m (No. 87), and is displaying in (Fig. 1). In this figure, the left side illustrates the total crystallographic structure; the right side shows only the magnetic structure atoms. The double perovskite $Pb_2FeReO_6$ exhibits a similar magnetic origin to that of $Sr_2FeMoO_6$ [35].

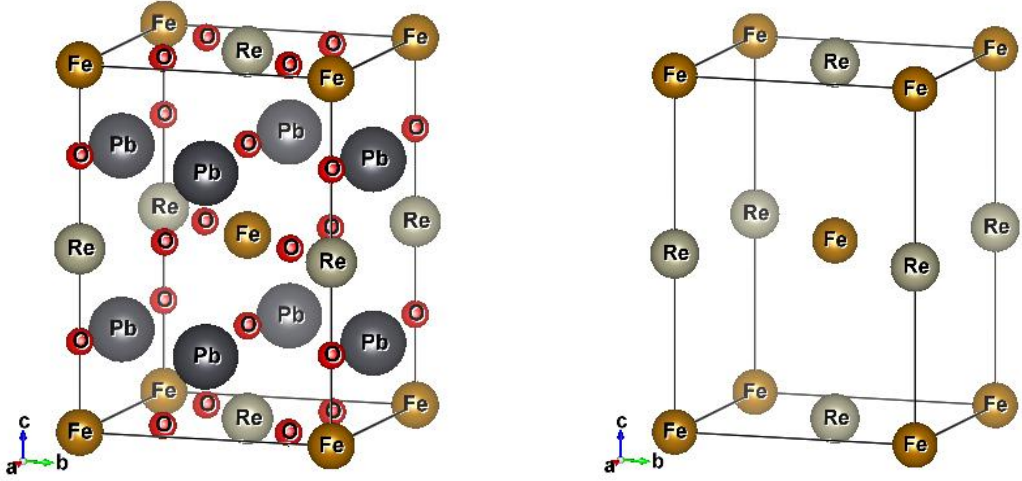

*Figure 1. The geometry of the Pb2FeReO6 using Vesta software [23]. In left side Crystallographic structure, in right side only magnetic structure.*

The Re sublattice has spins σ and takes the values ± 1 and 0, the Fe sublattice has spins S that can takes six values: ± 5/2, ± 3/2 and ± 1/2. Based on Ising model, the Hamiltonian controlling this system is as following:

$$\mathcal{H} = -J_1 \sum_{<i,j>} S_i S_j - J_2 \sum_{<k,l>} \sigma_k \sigma_l - J_3 \sum_{<m,n>} S_m \sigma_n - D_S \sum_i S_i^2 - D_\sigma \sum_i \sigma_i^2 - H(\sum_i (S_i + \sigma_i)) \qquad (1)$$

The notation <i, j>, <k, l> and <m, n> stand for the first nearest-neighbors spins. We performed and combined a MCS with Metropolis algorithm to simulate the magnetic behavior of this system.

J1: the interaction coupling between the nearest neighbors of Fe-Fe atoms (J1=0.89 eV [8]).

J2: the interaction coupling between the nearest neighbors of Re-Re atoms (J2=0.89 eV [8]).

J3: the interaction coupling between the nearest neighbors of Fe-Re atoms (J3=0.89 eV [8]).

$D_S$ and $D_\sigma$: represent the crystal fields of the two sub-lattices Iron and Rhenium, respectively. The origin of this field is the competition between Fe–O and Re–O interactions in the compound. For raison of simplicity, we make $D = D_S = D_\sigma$ (D=3 eV [8]). H is the external magnetic field.

*Table 1. Coordinates of $Pb_2FeReO_6$ [14].*

| Site | Location | X | Y | Z |
|------|----------|---|---|---|
| Pb | 8g | 0.000 | 0.500 | 0.2244 (1) |
| Fe1 | 2a | 0.000 | 0.000 | 0.000 |
| Fe2 | 2b | 0.000 | 0.000 | 0.500 |
| Re1 | 2a | 0.000 | 0.000 | 0.000 |
| Re2 | 2b | 0.000 | 0.000 | 0.500 |
| O1 | 8h | 0.246(15) | 0.252(11) | 0.000 |
| O2 | 4e | 0.000 | 0.000 | 0.247(17) |

## III. Monte Carlo Simulation

In this study, we investigate the magnetic properties of the perovskite $Pb_2FeReO_6$ using MCS. The magnetic structure is based on the position of the magnetic atoms, which are identified with Iron (Fe) and Rhenium (Re). This computation is founded on the Hamiltonian given in Eq. (1). For every spin configuration, $10^5$ Monte Carlo steps are performed. At each MCS step, all the sites in the system are visited and single-spin flip attempts are made. The flips are accepted or rejected according to the probability based on the Boltzmann statistics. The free boundary conditions are applied for the super cell lattice size 5x5x5.

At equilibrium, averages of the energy, the partial-total magnetizations and the partial-total magnetic susceptibility of the system have been calculated. We calculate, on each iteration, the internal energy per site (Eq.2). Afterward, we determinate the magnetization and we finally extract the magnetic susceptibility.

The internal energy:

$$E = \frac{<\mathcal{H}>}{N_T} \qquad (2)$$

Where $N_T = N_\sigma + N_S$ is the total numbers of spins in the super-cell unit, $N_\sigma$ being the number of $\sigma$ spins and $N_S$ is the total number of S spins.

The partial magnetizations are calculated as follows:

$$M_S = \langle \frac{1}{N_S} \sum_i S_i \rangle \tag{3}$$

$$M_\sigma = \langle \frac{1}{N_\sigma} \sum_i \sigma_i \rangle \tag{4}$$

The total magnetization is given by:

$$M_T = \frac{N_S.M_S + N_\sigma.M_\sigma}{N_T} \tag{5}$$

The partial and total susceptibilities are given by:

$$\chi_S = \frac{<M_S^2> - <M_S>^2}{K_B T} \tag{6}$$

$$\chi_\sigma = \frac{<M_\sigma^2> - <M_\sigma>^2}{K_B T} \tag{7}$$

$$\chi_T = \frac{<M_T^2> - <M_T>^2}{K_B T} \tag{8}$$

Where $K_B$ is the Boltzmann constant fixed at the value: $K_B=1$ and T being the absolute temperature.

## IV. Results and discussion

In this section, we analyze and discuss the ground states phase diagrams of the double perovskite $Pb_2FeReO_6$ at vanishing temperature (T = 0 K). Hence, we provide interpretations of the effect of physical parameters on the magnetizations and the susceptibilities.

### IV.1. Study of the ground state phases

In this part, we study the ground state phase diagrams indifferent plans in order to establish and discuss the all-possible configurations. Using the Hamiltonian of Eq.(1), we computed the energies of all possible 6x3=18 configurations. Figs. 2.a, …, 1.f show the obtained results in different planes corresponding to different physical parameters. Indeed, Fig. 2.a illustrates the twelve stable configurations plotted in the plane (J1, J2) for J3=-1, in the absence of both magnetic external and crystal fields (H=D=0). These twelve stable phases are: (±1/2, ±1), (±3/2,

±1), (±5/2, ±1). In fact, a perfect symmetry is present in this figure regarding the axis (J2=-5) when varying the parameter J1. For the specific point (J2=-5, J1=-5), we note the coexistence of all stable phases. This is due to the competition between the different exchange coupling interactions for the magnetic spins S=5/2 and σ=1.

In Fig. 2.b, plotted in the plane (J1, J3) for J2=-1, we found that the phases (±1/2, 0) and (±5/2, 0) are appearing for the crystal field value (D=0) and the magnetic external field (H=0). These phases are stable only for negative and large values of the parameter J3.

When fixing the interaction between Fe-Fe atoms (J1=-1), the Fig.2.c shows the corresponding stable configurations plotted in the plane (J2, J3). There is only six stable configurations, namely: (±5/2, 0) and (±5/2, ±1). When varying the exchange coupling interactionJ2 between Re-Re atoms, and keeping that one between Fe-Fe atoms at the fixed value of J1=-1, the extremely values of the magnetic moments of spin of Fe (S=±5/2) are found to be stable in this figure. Similarly, the stable phases shown in Fig.2.d are still found to keep the extremely values for the both spins S and σ, namely: ±5/2 and ±1, respectively.

In Fig. 2.d, plotted in the plane (H, J2) for J1=J3=-1 and D=0, we found that only four stable configurations are present in this figure, namely: (±5/2, ±1).A perfect symmetry is found in this figure regarding the axis H=0. Also, the sign of each stable configuration is the same as that one of the external magnetic field. This means that the stable configurations with positive values are found for positive H values and vice-versa.

To inspect the effect of both the magnetic external and the crystal fields on the stable configurations, we illustrate in Fig. 2.e plotted in the plane (H, D), the obtained results for the fixed values of J1=J2=J3=-1. From this figure, we note that the increasing crystal field values effect is to allow the apparition of the eight stable phases: (-5/2, -1), (5/2, 1), (-3/2, -1), (-1/2, -1) (-1/2, 0), (1/2, 0), (1/2, 1), (3/2, 1). A perfect symmetry is present in this figure, according to the axis H=0. Also, it is found that for large and positive values of the crystal field, the only stable phases are those corresponding to the extreme moment of spin values: ±5/2 and ±1, respectively.

To complete this ground state phase diagram discussion, we provide the stable phases in Fig. 2.f, plotted in the plane (D, J1) for H=0 and J2=J3=-1. From this figure, it is clear that the only stable phases are those already found in the other phase diagrams, namely: (±5/2, 0), (±1/2, 0), (-5/2, 1), (5/2, -1), (-1/2, 1) and (1/2, -1). It is worth to note that each configuration and its opposite are both stable in each region of this figure, see Fig. 2.e.

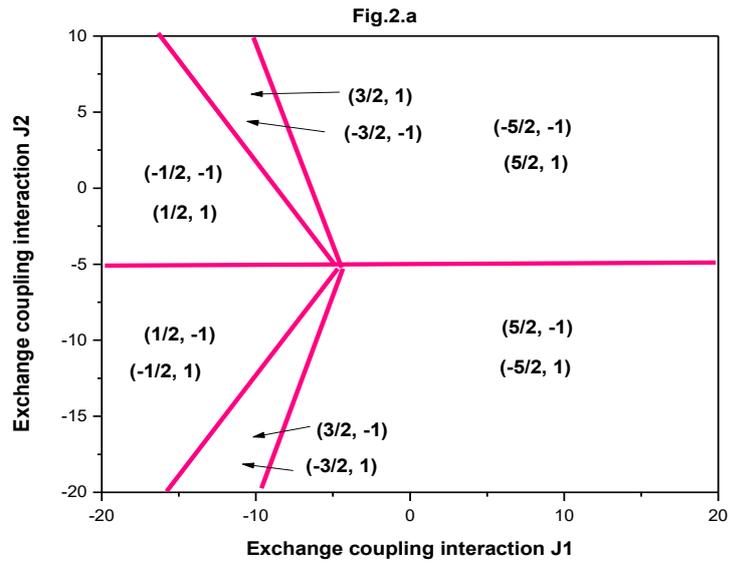

Fig.2.a

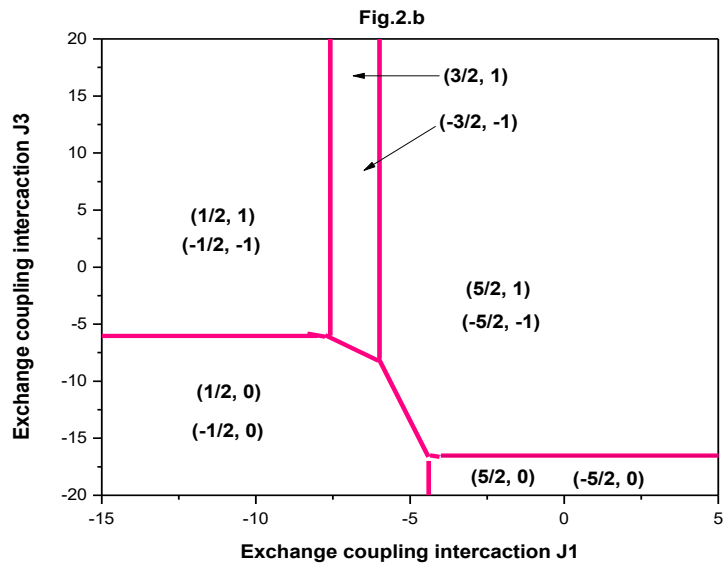

Fig.2.b

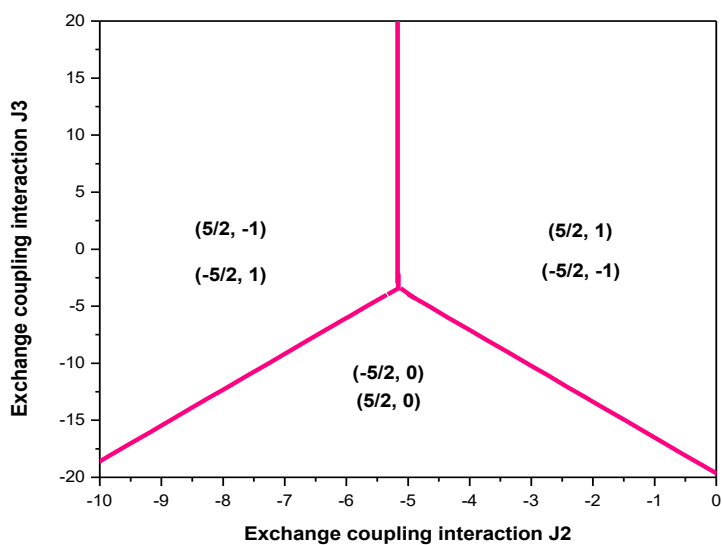

Fig.2.c

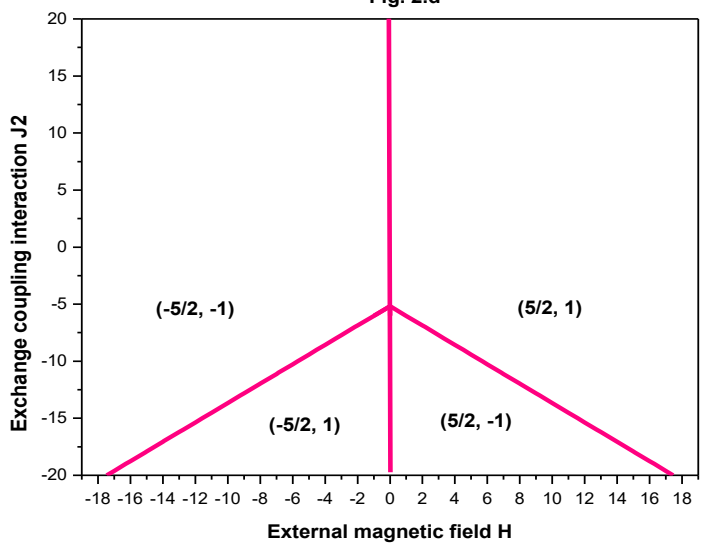

Fig. 2.d

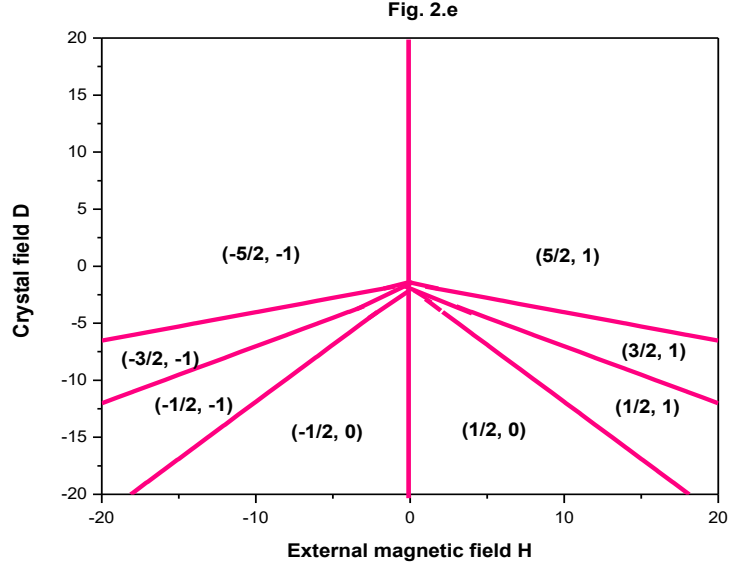

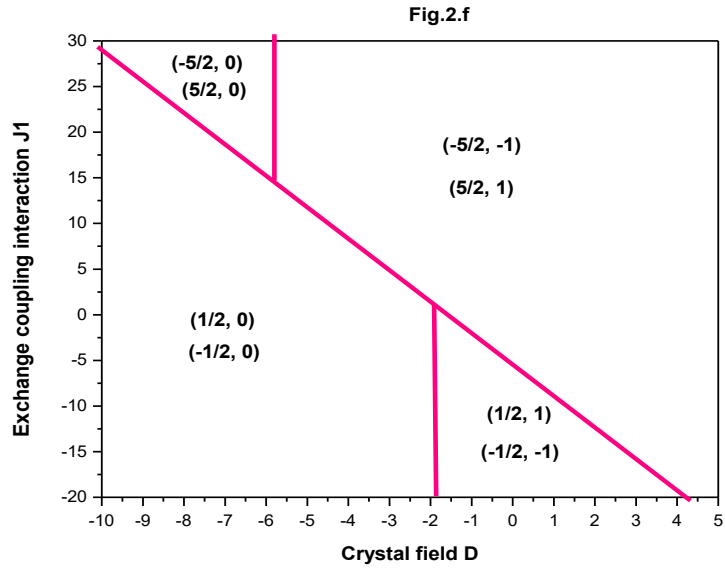

*Figure 2. Ground phase state diagrams of the compound $Pb_2FeReO_6$ in different planes: (a) in the plane (J1,J2) for J3=-1, in the absence of both magnetic external and crystal fields (H=D=0);(b) in the plane (J1,J3) for J2=-1 and H=D=0; (c) in the plane (J2,J3) for J1=-1 and H=D=0;(d) in the plane (H,J2) for J1=J3=-1 and D=0; (e) in the plane (H,D) J1=J2=J3=-1 and (f) in the plane(D,J1) for H=0 and J2=J3=-1.*

### IV.2. Monte Carlo Simulations

In order to study the critical properties of the compound $Pb_2FeReO_6$, we perform Monte Carlo simulations under Metropolis algorithm. In fact, such calculations are performed based on the Hamiltonian given in Eq. (1). At each spin configuration, a number of $10^5$ Monte Carlo steps are performed. When running a Monte Carlo simulation, we discard the first $10^4$ generated configurations. Single-spin flip attempts are made and the all sites of the system are visited. The flips are accepted or rejected according to the well-known probability of the Boltzmann statistics.

The obtained results are presented in Fig. 3, showing the total magnetization and susceptibility as a function of temperature for H=0, D=3, J1=J2=J3=0.89. From this figure, for low temperature values, the total magnetization reaches the exact value $M_T$=1.75, in good agreement with the value of the ground state which is $M_T$=(5/2+1)/2 for the fixed values D=3, J1=J2=J3=0.89. The total magnetization decreases when increasing the temperature. The total susceptibility reaches its maximum at the transition temperature corresponding to Tc=15 K, and changes its behavior above this point.

The behavior of the total magnetization as a function of crystal field is illustrated in Figs. 4.a, 4.b and 4.c, for H=0, selected values of the other physical parameters. In Fact, Fig. 4.a shows such behavior for selected temperature values: T=0.01, 5 and 10 K. From this figure, it is found that for low values of crystal field D ≤ -5, there are no crystal field effect on the attitude of the total magnetization. However, the effect of the crystal field in the region -5 ≤ D ≤ 5 is to decrease abruptly the total magnetization. Beyond the crystal field value D>5, the total magnetization reaches rapidly the saturation value $M_T$=1,75. The pertinent parameter here is the crystal field effect since the changing sign of this parameter leads to a brutal change of the total magnetization behavior. The effect of varying the exchange coupling interaction between J3 between Re-Fe atoms is illustrated in Fig. 4.b, for the values J3=+3, -3 and -1. It is found that the total magnetizations are strongly affected by the J3 variations in the interval of the crystal field -5<D<+5. Outside this interval, for high values of the crystal field, the total magnetizations are not affected by the variations of the parameter J3, see Fig. 4.b.

When varying the exchange coupling interaction between J1 between Fe-Fe atoms, the behavior of the total magnetizations is presented in Fig. 4.c for the specific values: J1=+3, -3 and -1. Similarly to Fig. 4.b, outside the interval -5<D<+5, the total magnetizations are slightly affected by the variations of the parameter J1.

The behavior of the total magnetization as a function of the magnetic external field is illustrated in Figs. 5.a,…, 5.d. In all the following figures, the magnetization behavior of the antiferromagnetic and the ferromagnetic films causes a shift in the soft magnetization curve, this phenomena is due to the exchange bias effect, see for example Refs. [36-38]. In Fig. 5.a, we present such behavior for different temperature values: T=0.01, 5 and 10 K. As it is expected, the surface of the hysteresis loops decreases when increasing the temperature values. The corresponding coercive field also decreases when the temperature increases. The effect of varying the crystal field when increasing the external magnetic field is illustrated in Fig. 5.b, for D=+3, -3 and 0. Contrary to the temperature effect, the effect of increasing the crystal field is to increase both the surface and the corresponding coercive field of the loops.

To inspect the effect of varying both the parameters J1 and J3 are presented in Fig. 5.c and Fig. 5.d, respectively. In fact, Fig. 5.c is devoted to the variation of the total magnetization as a function of the magnetic external field for J1=+1, -1 and -3. The hysteresis surface loops are practically not affected by the increasing values of the exchange coupling interaction between Fe-Fe atoms J1.

In Fig. 5.d, plotted for J3=+1, -1 and -3, we show the effect of varying the exchange coupling interaction between Re-Fe atoms on the hysteresis loops. For 3=+1, there is apparition of intermediate steps, probably due to the low interaction between Re-Fe atoms.

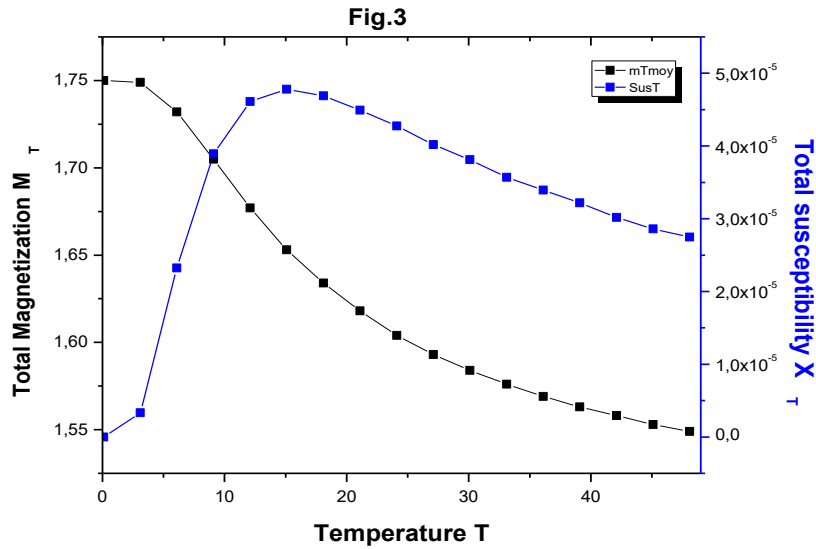

*Figure 3. Total Magnetization and total susceptibility as a function of temperature for H=0, D=3 and J1=J2=J3=0.89.*

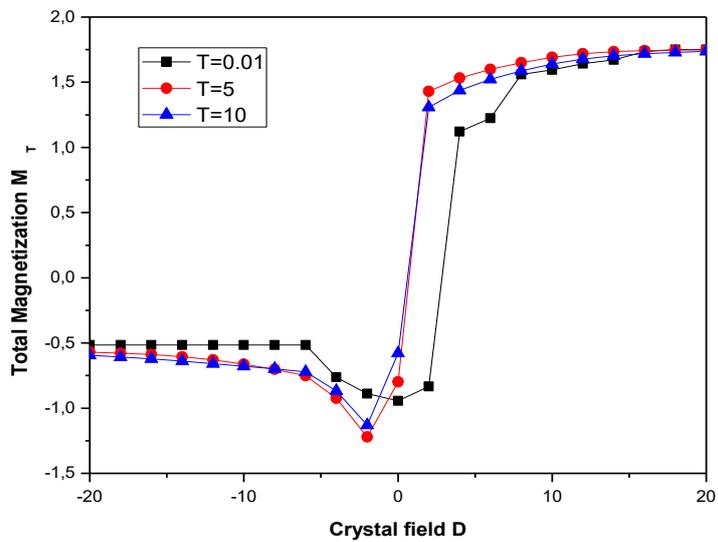

**Fig.4.a**

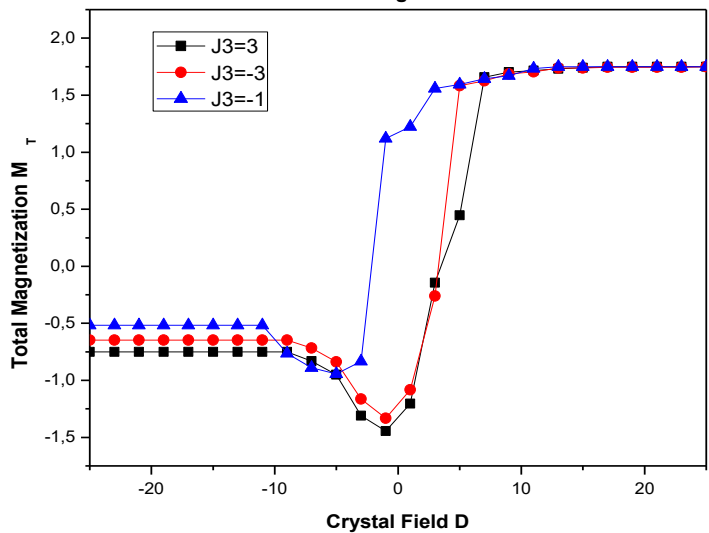

**Fig.4.b**

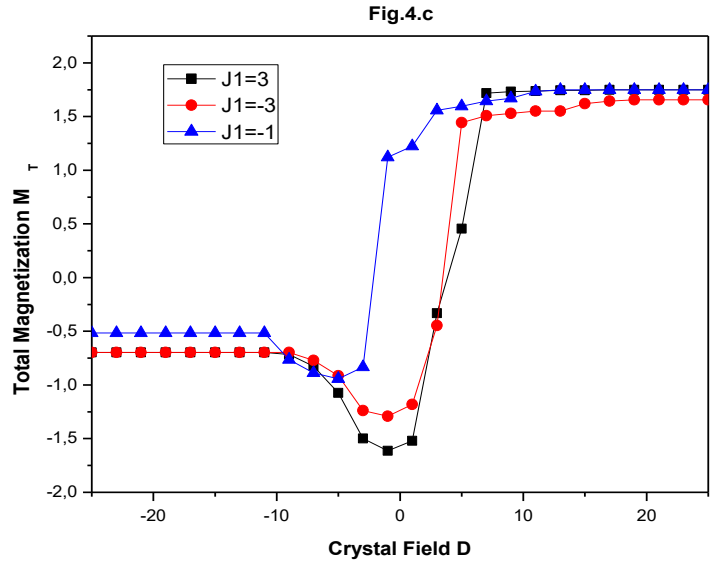

*Figure 4. Total magnetization as a function of crystal field for H=0: in (a) for J1=J2=J3=-1, and T=0.01, 5 and 10 K; in (b) for T=0.01 K, J1=J2=-1 and J3=+3, -3 and -1; in (c) for T=0.01 K, J2=J3=-1 and J1=+3, -3 and -1.*

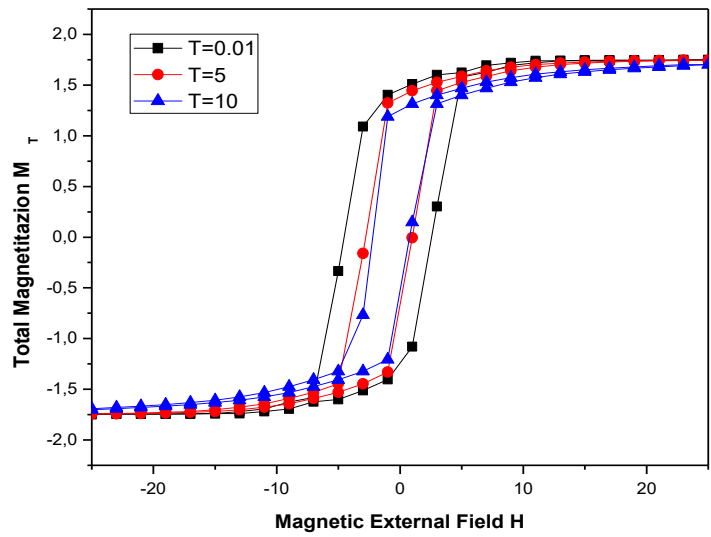

**Fig.5.a**

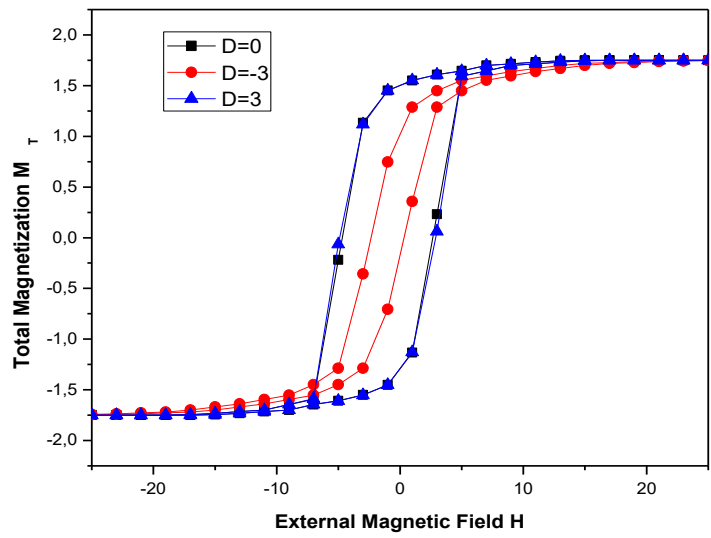

**Fig.5.b**

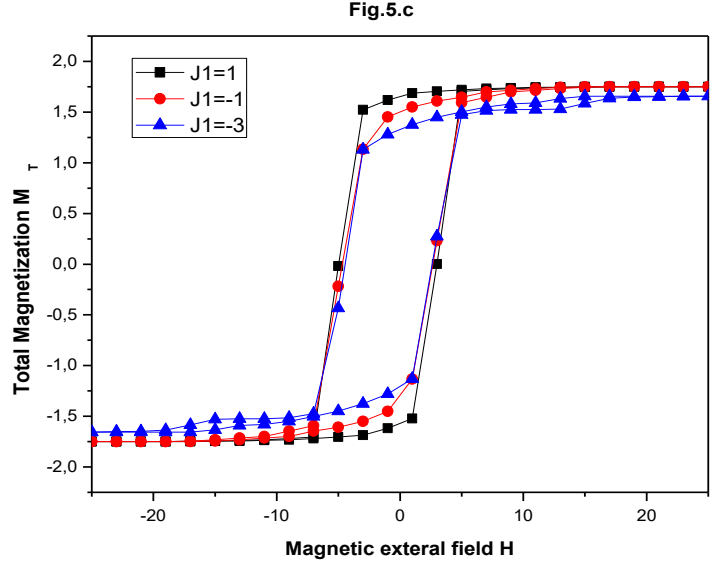

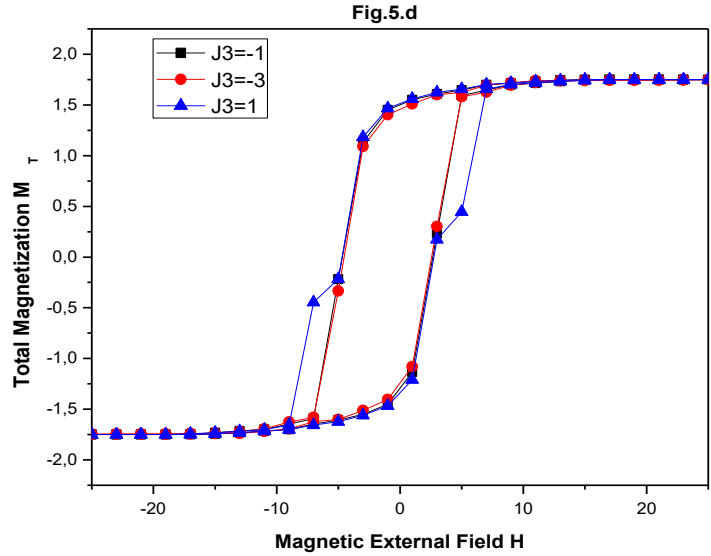

*Figure 5. Total Magnetization vs. magnetic external field: in (a) for J1=J2=J3=-1, D=0 and T=0.01K, 5 and 10 K; in (b) for J1=J2=J3=-1, T=0.01 and D=+3, -3 and 0; in (c) for J2=J3=-1, D=0, T=0.01K and J1=+1, -1 and -3; in (d) for J1=J2=-1, D=0, T=0.01K and J3=+1, -1 and -3.*

## V. Conclusion

Using the Monte Carlo simulations (MCS) we have inspected and predicted the phase diagrams and the magnetic properties of the half-metallic double perovskite $Pb_2FeReO_6$. To define the stable states between the 6x3=18 possible configurations, based on the Hamiltonian of this system in the absence of any temperature, we calculated the energy corresponding to every configuration and taken the minimum one. The more stable configurations are; (-5/1, -1) and (5/2, 1), they coexist in all planes. On the other hand, we have examined the behavior of the total magnetizations and the total susceptibilities as a function of the temperature, the crystal field, the exchange coupling interactions and the external magnetic field. To complete this study, we analyzed and discussed the hysteresis loops for fixed values of temperature and the other physical parameters. It is found that for positive values of the exchange coupling interaction J1 between Fe-Fe atoms, the maximum value of the total magnetization ($M_T$=1.75) is reached rapidly. On the other hand, we found that increasing the crystal field and the exchange coupling interaction values lead to increase the surface, the corresponding coercive field and the remanent magnetization of the loops. Moreover, by increasing the temperature, the surface, the coercive field and the remanent magnetization of the loops decrease. The temperature of transition is found to be about $T_C$=11 k for the fixed physical parameters.